# Focusing Optical Phased Array for Optically Enabled Probing of the Retina with Subcellular Resolution

Pedram Hosseini,[1] Prachi Agrawal,[1] Alireza Tabatabaei Mashayekh,[1] Sandra Johnen,[2] Jeremy Witzens[1] and Florian Merget[1]

[1] Institute of Integrated Photonics, RWTH Aachen University, 52074 Aachen, Germany
[2] Department of Ophthalmology, University Hospital RWTH Aachen, 52074 Aachen, Germany
phosseini@iph.rwth-aachen.de

**Abstract.** We present a silicon-nitride-based optical phased array with built-in focusing and steering capability, that operates at 522 nm and is aimed at complementing a micro-electrode array for joint electrical and optical probing of retinal tissue. It achieves subcellular resolution with a beam diameter of 1.4 µm at a focal point located above the chip. Targeted cellular excitation can be achieved by steering the beam through a combination of wavelength tuning and simplified thermo-optical phase shifters with a single electrical input for each of transverse beam steering and selection of the focal plane.

## 1      Introduction

*Ex-vivo* recording of electrical retinal activity performed with microelectrode arrays (MEAs) and time-varying blanket optical illumination [1] can be augmented by spatially selective optical excitation. The ability to target single photoreceptor cells appears a desirable characteristic for light delivery systems, together with the ability to span large fields of view (FOV) covering a substantial portion of the retina. The joint or sequenced excitation of selected photoreceptor cells would also be a powerful tool to gain new biological insights.

State-of-the-art approaches like micro-LEDs [2, 3] and optical holographic setups [4] have limitations, such as the dissipation of heat generated by micro-LEDs into the tissue or limited resolution and FOV. Moreover, focusing light onto the tissue on a micrometer scale with objective lenses results in bulky free space setups. Using a photonic integrated circuit (PIC) as a substrate for the MEA enables the co-integration of programmable optical emitters with a small footprint [5]. Beam shaping and steering can be obtained at the chip scale with optical phased arrays (OPAs) [6]. Even though OPAs have been intensely studied at longer wavelengths for free-space optical communication or light detection and ranging (LiDAR) [7, 8], there has been more limited research in the visible spectrum, which is more applicable to biophotonics applications such as optogenetics [9, 10] or flow cytometry [11]. The fabrication of PICs on silicon substrates with established industrial CMOS-manufacturing processes reduces the fabrication cost and process variability. The large transparency window and low losses of silicon nitride (SiN) technology makes it ideal for short wavelength applications [12, 13].

Surface emitting OPAs are promising integrated devices for replacing free-space beam shapers, that enable beam scanning [8] as well as advanced beam shaping, such as focusing [14, 15], multi-beam emission or patterned illumination, by emitting light from an array of emitters with programmable phases. These can be statically set, tuned with thermo-optic or micro-mechanical phase shifters, or set by optical delay lines and tuning of the light source wavelength [16]. Two-dimensional (2D) OPAs require a complex actuation system and a large number of control signals for phase tuning. Reducing the complexity of such systems is an active field of research [17]. For system simplification, using the wavelength dependency of the grating coupler (GC) emission angle can be used to steer the beam in one direction. Using a one-dimensional OPA with a combination of wavelength tuning and thermo-optical phase shifters leads to 2D beam steering with considerably less complexity [18].

Here, we present a one-dimensional (1D) phased array with built-in focusing capability fabricated in the SiN PIC platform (Section 2). Light can be steered by 3 degrees in the longitudinal direction by means of input wavelength tuning in a range from 510 nm to 530 nm. In the transverse direction, along which the GCs are arranged, focusing is achieved above the chip, with good consistency with the simulation results. In Section 3, we describe a thermal tuner concept with supporting simulations, by which beam steering in the transverse direction and variation of the focal spot position can be each achieved with a single tuning signal.

## 2   Focusing One-Dimensional Optical Phased Array

### 2.1   Design

Fig. 1(a) describes the architecture of the SiN-PIC-based focusing OPA. It is implemented in the double-stripe SiN waveguide platform from LioniX International BV with layer thicknesses optimized for operation in the visible range, deposited on an 8 µm buried oxide (BOX) layer [13, 19]. It consists of a core with a 50 nm silicon dioxide ($SiO_2$) layer sandwiched between two SiN layers that are 27 nm (bottom) and 80 nm (top) thick. Finally, the core is encapsulated within a 4 µm thick top cladding $SiO_2$ layer over which metal heaters are fabricated for phase tuning. The PIC comprises an edge coupler for light in-coupling from a laser source with a 520 nm central wavelength. The light is evenly distributed by a 1×16 passive light distribution network that employs 4 levels of identical 1×2 multi-mode interferometers (MMIs). These have a length of 15.3 µm, a width of 3 µm, and an access waveguide width of 0.8 µm (Fig. 1 (c)). At the output of the MMI, the waveguide centers are offset by ±0.75 µm relative to the center of the MMI. For optical focusing, delay lines of varying length are implemented in the resulting waveguide array (Fig. 1(a)), that introduce a phase delay depending on the waveguide index in a parabolic manner (Fig. 1(b)). The GCs, with a periodicity of 600 nm and a fill factor of 0.5, are uniformly spaced and arranged with a 2.4 µm pitch (Fig.

1 (d)). Each grating emitter has a 2 μm width and 20 trenches that are etched in the top SiN layer only. A 10-μm long adiabatic taper prevents higher-order mode excitation while transitioning from a single-mode waveguide (0.4 μm, fully etched, with an effective index $n_{eff} = 1.68$) to the multi-mode waveguide width at the grating section. The GCs are simulated to scatter 51% of the light towards the top of the chip along an angle oriented 44° from the surface normal.

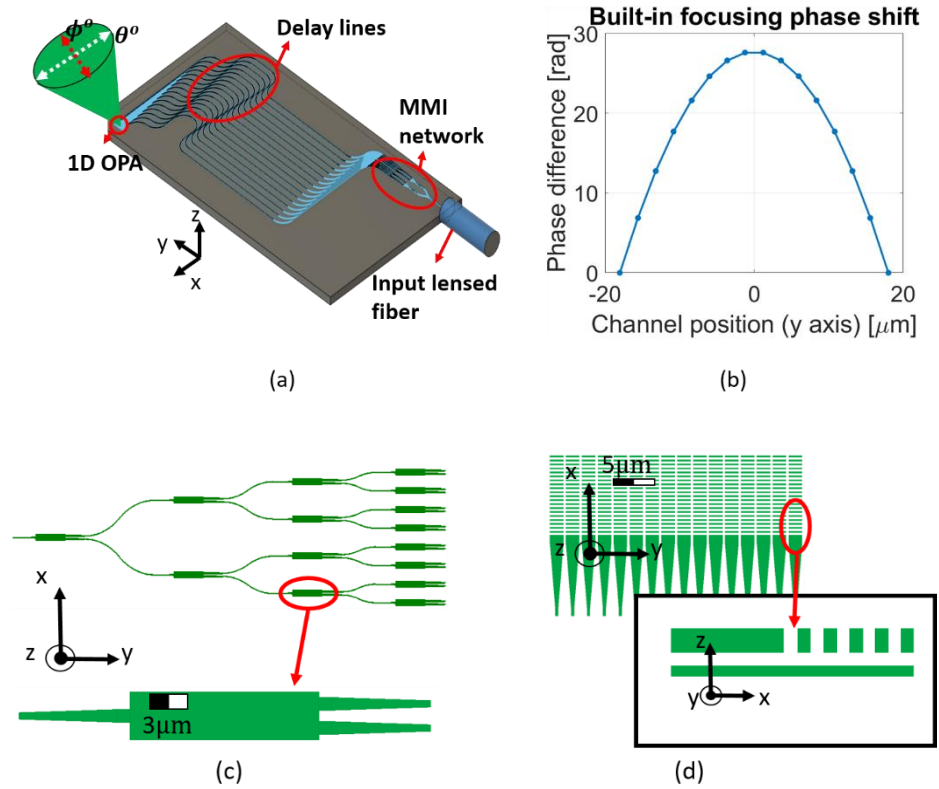

**Fig. 1.** Schematic of the designed 1D OPA with built-in focusing. It includes an edge coupler for in-coupling of the light, a 1x16 passive light distribution network based on 1x2 MMIs, delay lines for generating the phase delays required for focusing the light above the chip, and optical GC emitters. $\theta$ and $\phi$ describe the beam propagation in the longitudinal and transverse directions, respectively. (b) Phase delay induced in each of the OPA waveguide channels. (c) Layout of the passive light splitter network based on 1x2 MMIs. (d) Top and side views of the 1D OPA layout (including the layer stack of the double stripe SiN platform).

## 2.2    Simulation

Fig. 2 depicts the simulated far-field emission of the OPA. The near-field of a single GC is first simulated and recorded by utilizing the 3D finite-difference time-domain (FDTD) method. The field emitted at the surface of the PIC by the 1×16 GC array is then generated from it, by tiling the GC emission after applying the expected emission phase. A far-field transform is then applied in Fourier space to predict the emitted field profile at different heights above the chip. Three beams are generated with a spatial separation of 12.2° consistent with the GC pitch and a spacing of 15.3 µm from each other at the focal plane. The center beam and each of the two side-beams carry 38% and 26% of the total, top-emitted power, respectively. As expected from the built-in delay lines, the main beam focuses at a height of 50 µm above the surface of the chip (measured along z), that allows to accommodate the thickness of a biological sample, and propagates relative to the z-direction with the GC propagation angle of 44°, as depicted in Fig. 2(a). Fig. 2(b) shows the illumination pattern along the $(y,u_z)$ plane oriented along the direction of propagation of the beam $(u_z)$ and Fig. 2(c) shows the illumination pattern in the $(u_x,y)$ plane perpendicular to its direction of propagation, at the focal point of the beam. The full width at half maximum of the beam (FWHM), recorded along the y-direction along which it is focused, is 1.4 µm and is small enough to target a single cell. Along $u_x$ the beam forms an extended sheet with a FWHM equal to 6.62 µm, and the FWHM of the beam along its direction of propagation $(u_z)$ is 23 µm.

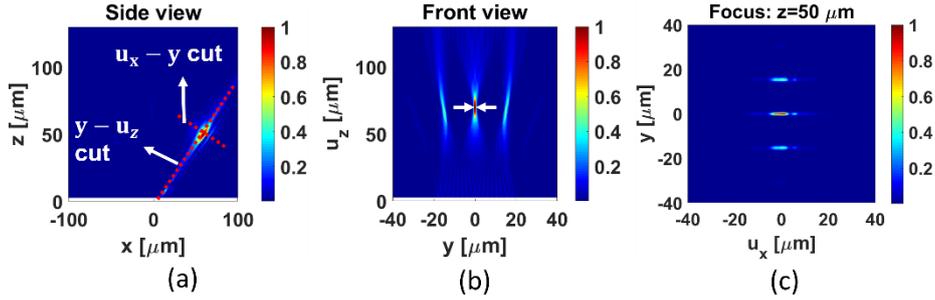

**Fig. 2.** Simulated field intensities resulting from 3D FDTD followed by a far-field projection at wavelength 520 nm. (a) Side view of the illumination pattern revealing the GC emission angle 44°; illumination patterns along (b) the $(y,u_z)$ plane along the direction of propagation of the beam and (c) the $(u_x,y)$ plane perpendicular to its direction of propagation.

### 2.3 Measurement and optical characterization

To characterize the OPA, we orient a CMOS camera equipped with an objective lens such that its optical axis coincides with the direction of propagation of the central beam. The position of the camera is then varied to record images of the out-coupled light at different distances above the surface of the PIC. Fig. 3(a) depicts the field intensity recorded at a height at which the three beams reach their focal points. It reveals a FWHM in the direction of focusing (y) of less than 2 µm, limited by the resolution of the camera. The FWHM recorded along $u_x$ is 7 µm, in good agreement with simulations. There is also a good consistency between the measured positions of the side lobes, at ±15.5 µm from the central beam, and the simulations.

Several such images are then recorded at different camera heights and stitched together to generate the intensity profile in the $(y,u_z)$ plane shown in Fig. 3(b). In the (x,z) plane, the beam propagates with an angle of 38° relative to the surface normal. The 6° difference in the propagation angle is likely caused by fabrication tolerances or uncertainties in the measurement setup resulting from the alignment of the CMOS camera.

Wavelength dependent beam steering in the longitudinal direction was also measured in the wavelength range 510 nm - 530 nm and was found to be 3 degrees, in good agreement with simulations.

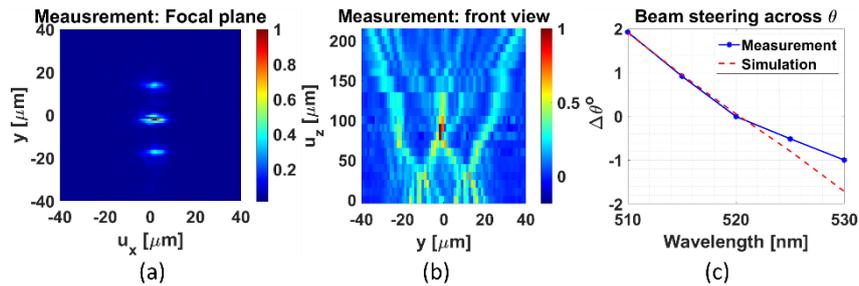

**Fig. 3.** Beams imaged with the optical axis of the camera aligned in the direction of propagation of the central beam. In (a), the camera was positioned such that the beam profiles in the focal plane of the PIC are projected onto the imager by its objective. Several such recordings are taken at different camera heights, recentered and stitched together to obtain the front view profile shown in (b). (c) Steering of the beams in the longitudinal direction ($\theta$) as a function of wavelength, relative to the angle of propagation at 520 nm. (a) and (b) were recorded at 522 nm.

## 3  Thermo-optical phase shifter for beam steering

Incorporating thermo-optical phase shifters into the PIC enables the application of a tunable linear phase shift across the OPA and thus transverse beam steering ($\phi$) in addition to the longitudinal beam steering ($\theta$) achieved through wavelength tuning. However, equipping each channel with a separate heater requires a large number of control signals and significantly enlarges the device footprint once the required pads are included.

Consequently, adopting a single heating element capable of inducing a linear phase shift across the waveguide array with a reduced number of control inputs significantly increases the practicability of the system [20]. Fig. 4(a) shows a meandered heater configuration that allows applying a linear phase shift across the waveguide array with a single electrical current and a single pair of pads. To steer the light in both directions, that require linear phase shifts with increments of opposite signs, two such meandered heaters are implemented, with the top one decreasing the phase shift and the bottom one increasing the phase shift as one moves down the waveguide array.

These meandered phase tuners are simulated in a 3D model considering the 4.3 Ω/sq sheet resistance of the heater layer, resulting in a 270 Ω resistance for their 20 μm width and 1.25 mm length. Cascading several such heater elements along the length of the waveguide array, larger cumulative deflections can be achieved. Considering two cascaded elements and applying a 55-mA current in each, the beam can be deflected by 3 degrees taking the thermo-optic coefficients of SiN and $SiO_2$ into account, see Fig. 4(b).

A similar concept can be used to apply a tunable parabolic phase shift across the waveguide array, in addition to the one that is already passively set, in order to tune the height at which the beam reaches its focal spot (Fig. 5(a)). This meandered heater has a length of 1.12 mm and a resistance of 241 Ω. Cascading again two such elements and applying a 50 mA current to each shifts the focal spot from the passively built-in 50 μm height (measured along the z-direction) to 38 μm. Considering the direction of propagation of the beam, this corresponds to a displacement of 16.7 μm along $u_z$, which substantially exceeds the half width at half maximum of the beam in that direction. Adding meandered heaters with the opposite convexity would allow tuning the focal spot position in both directions.

Experimental results for these meandered heaters will be reported in a later publication.

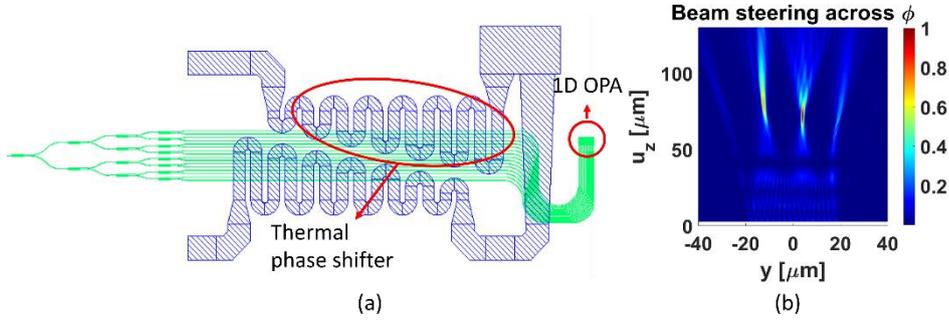

**Fig. 4.** (a) Meandered phase tuner design for linear phase shifting. (b) Simulation result after applying 55 mA current over two cascaded 1.55 mm long meandered heaters.

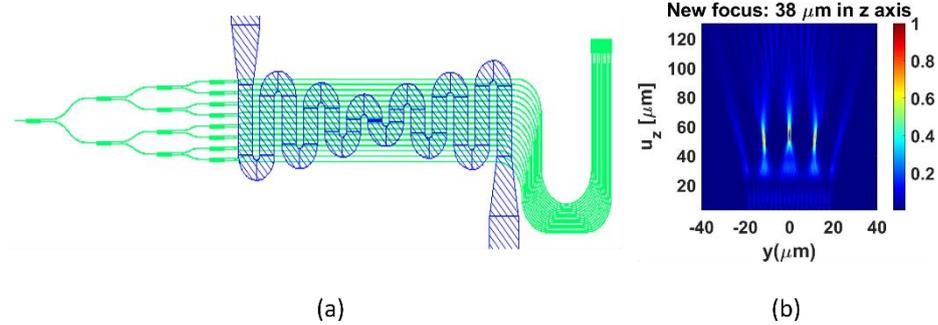

**Fig. 5.** (a) Meandered phase tuner design for shifting of the focal plane. (b) Simulation result after applying 50 mA current over two cascaded 1.12 mm long meandered heaters.

## 4   Conclusion

In this work, we experimentally demonstrate a focusing 1D optical phased array enabling the optical excitation of single cells located above the surface of the chip, with a spot FWHM of ~1.4 µm in the transverse direction along which the grating emitters are arranged. The PIC is implemented in SiN technology and aimed at simultaneous optical excitation of single photoreceptor cells and electrical recording in ex-vivo setups for the analysis of retinal activity. Three degrees of beam steering are achieved in the longitudinal direction by tuning the input wavelength between 510 nm and 530 nm, enabling targeted cell addressability. A concept for compact and easy to handle thermal actuation of beam steering in the transverse direction and shifting of the focal plane position was simulated. Meandered heater configurations allow the implementation of phase profiles of predetermined shape and tunable overall amplitude across the OPA, allowing steering or variable focusing with a single control signal. This significantly decreases the OPA's complexity, while enabling steering of the interrogation point in all three directions.

**Acknowledgements** Funded by the Deutsche Forschungsgemeinschaft (DFG) under Research Training Network 2610 "InnoRetVision," project B3.